\documentstyle[emulateapj,psfig]{article}
\lefthead{COLE, WOOD, \& NORDSIECK}
\righthead{MODEL POLARIMETRY}

\begin{document}


\title{Ultraviolet Imaging Polarimetry of the Large Magellanic Cloud.
II. Models}
 

\author{Andrew A. Cole,\altaffilmark{1,4} Kenneth Wood,\altaffilmark{2}
\& Kenneth H. Nordsieck\altaffilmark{1,3}}
\altaffiltext{1}{{\it cole@astro.wisc.edu}; Department of Astronomy,
	University of Wisconsin-Madison, 475 North Charter St., 5534
	Sterling Hall, Madison, WI, 53706.}
\altaffiltext{2}{{\it kenny@claymore.harvard.edu}; Smithsonian Astrophysical
	Observatory, 60 Garden Street, Cambridge, MA, 02138.}
\altaffiltext{3}{{\it khn@sal.wisc.edu}; Space Astronomy Laboratory, 
	University of Wisconsin-Madison, 1150 University Avenue, 
	Madison, WI, 53706.}
\altaffiltext{4}{{\it Current address:} Department of Physics \& Astronomy, University of
	Massachusetts, Amherst, MA 01003-4525.}

\setcounter{footnote}{0}

\begin{abstract}
Motivated by new sounding-rocket wide-field polarimetric images of the 
Large Magellanic Cloud (\cite{col99a}), we have used a 
three-dimensional Monte Carlo radiation transfer code to 
investigate the escape of near-ultraviolet photons from young
stellar associations embedded within a disk of dusty material
(i.e., a galaxy).  As photons propagate through the disk,
they may be scattered or absorbed by dust. Scattered photons
are polarized and tracked until they escape to be observed;
absorbed photons heat the dust, which radiates isotropically
in the far-infrared, where the galaxy is optically thin.
The code produces four output images: near-UV and far-IR flux,
and near-UV images in the linear Stokes parameters Q and U.  From these images
we construct simulated UV polarization maps of the LMC.  We
use these maps to place constraints on the star$+$dust geometry
of the LMC and the optical properties of its dust grains.
By tuning the model input parameters to produce maps that
match the observed polarization maps, we derive information about the
inclination of the LMC disk to the plane of the sky, and about
the scattering phase function $g$.  We compute a grid of models
with $i$ = 28$\arcdeg$, 36$\arcdeg$, and 45$\arcdeg$, and $g$ = 0.64,
0.70, 0.77, 0.83, and 0.90.   The model which best reproduces the
observed polarization maps has $i$ = 36$\arcdeg^{+2}_{-5}$ and 
$g$ $\approx$ 0.7.  Because of the low signal-to-noise in the data,
we cannot place firm constraints on the value of $g$.
The highly inclined models do not match the observed centro-symmetric
polarization patterns around bright OB associations, or the 
distribution of polarization values. 
Our models approximately reproduce the observed ultraviolet
photopolarimetry of the western side of the LMC; however, the
output images depend on many input parameters and are nonunique.
We discuss some of the limitations
of the models and outline future steps to be taken; our models make
some predictions regarding the polarization properties of diffuse
light across the rest of the LMC.
\end{abstract}

\keywords{polarization --- methods: numerical --- galaxies: individual (LMC)
	--- ISM: dust, structure}

\section{Introduction}

Polarimetric imaging provides a unique window on the 3-dimensional
structure of astrophysical objects, and therefore on the physical
processes operating in a wide range of stellar and interstellar 
environments.  The most important processes giving rise to interstellar 
polarization are scattering by dust, and transmission through aligned
dust grains.

Imaging polarimetry has been applied to many targets as a 
primary means of determining the scattering properties of dust,
and to obtain geometric information on extended and complex sources
(e.g., reflection nebulae, active galactic nuclei, and comets).

The vacuum ultraviolet is an especially favorable wavelength 
regime for these studies; polarimetric efficiencies are high, and
polarized backgrounds are low (\cite{nor93}).  Moreover, a relatively small
number of bright stars emit the majority of VUV photons, greatly
simplifying the accurate tracing of source-scatterer-detector
geometry over the case in the optical and near infrared.

The University of Wisconsin's {\it Wide-Field Imaging Survey 
Polarimeter} (WISP) was developed to obtain the first wide-field
astronomical polarization images in the vacuum ultraviolet.  
This rocket-borne instrument has been flown three times to date,
providing high-quality images of reflection nebulosity in the 
Pleiades open cluster (\cite{gib95}, \cite{gib97}, Gibson 
\& Nordsieck 1999, in preparation), and 
photopolarimetry of Comet Hale-Bopp (\cite{har97b}).  Additionally,
WISP obtained polarimetric images of the Large Magellanic Cloud (LMC);
these observations represent the first wide-field, UV polarization images ever
obtained (\cite{nor96}).  Analysis of the LMC data (\cite{col99a},
hereafter Paper I)
found that the diffuse UV light is polarized at a 5--10\% level,
consistent with starlight scattered by dust; that the strongest
source of illumination in the WISP field is the H {\small II}
complex N11; and that the UV starlight must account for most
of the heating of diffuse dust in the LMC.

In this paper, we report on our program to model the WISP 
polarization maps of the LMC using
a Monte Carlo radiation transfer code to constrain the 
optical properties and scattering geometry of the dust in
the LMC's diffuse interstellar medium.  \S1.1 and \S1.2, respectively,
describe the observational results which motivate this work,
and the interpretive issues addressed by our models.

In \S2, we describe in detail the basic astrophysical ingredients
of our models, which are the distributions in space, size, and 
luminosity of the illuminating OB associations and scattering
dust medium.  In \S3 we discuss the scattering and polarizing properties
of the dust grains and the way in which we parameterize the total
amount of dust present.  \S4 describes the Monte Carlo engine of
our radiation transfer code, which is innovative in its ability
to track the processing of near-UV photons into thermal IR radiation
by heated interstellar dust grains.
In \S5 the modelling algorithm is explained, relating the procedure
by which we explored parameter space for the ``best'' models.
\S6 presents the model images and polarization maps and the conclusions
we can draw about the inclination of the LMC disk and the scattering
asymmetry of its dust grains.  We are careful to note the many 
shortcomings of this simple model, which nonetheless reproduces
many of the observed near-ultraviolet and far-infrared properties
of the LMC for a reasonable set of inputs.

\subsection{Observations: the Wide-Field Imaging Survey Polarimeter}

A 1$\fdg$5 $\times$ 4$\fdg$8 area of the western side of the LMC
was observed with the rocket-borne Wide-Field Imaging Survey 
Polarimeter (WISP) on November 20, 1995.  4 $\times$ 80 second
exposures in an intermediate-band, near ultraviolet filter 
($\lambda$ = 2150 \AA, $\Delta \lambda$ = 300 \AA) were used
to create intensity and polarization maps of the field.
The observations were centered at $\alpha$ = 
04$\mathrm ^{h}$ 59$\mathrm ^{m}$, $\delta$ $-$67$\arcdeg$ 53$\arcmin$
$ $ (J2000.0) and aligned roughly north-south. The WISP
instrument is described in detail in \cite{nor93}; the reduction,
calibration, and analysis of the LMC flight data are given in
Paper I.

The minimum diffuse UV surface brightness, 
5.6 $\pm$3.1 $\times$ 10$^{-8}$ erg s$^{-1}$
cm$^{-2}$ \AA$^{-1}$ Sr$^{-1}$, is larger than any known stray light background,
and is clearly due to light originating within the LMC.  The surface
brightness of this diffuse UV background is correlated with areas of high
H {\small I} column density and is linearly polarized at the $\sim$10\% 
level.  This suggests that reflected OB starlight contributes at least half
of the LMC's diffuse UV background.  The ISM of the Large Magellanic Cloud
apparently acts as a kiloparsec-scale reflection nebula in the near ultraviolet.
Paper I found evidence for weak centro-symmetric scattering halos
around some of the large OB complexes in the WISP field.  The B2 complex
(Martin {\it et al.} 1976), however, lacked such a halo; this was 
interpreted to mean that B2 is located either within an H {\small I} 
hole or well above the plane of the LMC disk.

\subsection{Modelling Goals}

It is desirable to test the interpretation of Paper I;
to this end we have undertaken to model the radiation transfer
of ultraviolet photons from their origins in hot stars, through
the dusty ISM of the LMC's disk, to Earth.  Using these models
we hope to determine whether or not the observed level of
polarization is consistent with the reflection nebula interpretation.
We also wish to determine the expected polarization pattern around
B2 for a location within the disk; perhaps a non-detection of 
centro-symmetry is to be expected for this region.

Using a specialized Monte Carlo radiation transfer code, we 
set out to determine whether or not reasonable values for
the dust optical depth, scattering geometry, and dust grain optical
properties can account for the WISP observations.  Under the assumption
that the reflection nebula interpretation is correct, we can use the
polarization properties of the model to place constraints on the
dust properties and inclination of the disk of the LMC.

\section{Model Ingredients and Assumptions}

\subsection{Illuminating Sources}

The perfect model of the Large Magellanic Cloud would incorporate
the luminosity contributions of every field star and star cluster
into its input parameters.  This is obviously impractical, and so
we must find a more tractable subset of objects with which to 
illuminate the LMC's dust.  The star-formation rate of LMC field
stars has been roughly constant for the past $\approx$1--2 Gyr
(e.g., \cite{gal98}; \cite{wes97} and references therein).  This
recent activity has been accompanied by the formation of a large
number of ``blue populous'' star clusters and OB associations;
the young clusters of the LMC are both more frequent per unit
field star mass and individually larger than their Milky Way 
counterparts (e.g., \cite{els85}; \cite{van84}). 

Data from the
UIT instrument suggests that $\approx$75\% of the flux from the
LMC at $\lambda$ = 1500 \AA$ $ originates from stellar associations
within the regions of nebulosity catalogued by \cite{dav76} 
(\cite{par98}).  For the western side of the LMC (observed by
WISP), this interpretation holds true at 2150 \AA.  In the WISP
image, most of the well-detected sources can be identified with
OB associations (\cite{luc70}), or open clusters younger than
$\approx$200 Myr.  Clusters older than this, e.g., the massive
1 Gyr-old young globular cluster NGC 1783, are undetected in 
our image.  Individual supergiants among the field stars 
(\cite{san69}) can be detected, but are minor contributors to 
the total observed flux.  

We therefore chose to take as our illuminators the 122 OB associations of
\cite{luc70}, because a homogeneous dataset of ultraviolet photometry
at two wavelengths exists for the entire sample (\cite{smi87}, hereinafter
SCH).  Due to the lack of a uniform sample of ultraviolet photometry,
we have ignored the young open clusters in this first model; some
of these clusters, e.g., NGC 1818, NGC 1755, and NGC 1711, contribute
significant UV flux to the WISP image.

The positions of the OB associations from \cite{luc70} 
were transformed onto the model's rectilinear coordinate system
at a scale of 15$\arcmin$ per grid unit.  The scale was chosen
in order to accomodate output images of the entire LMC, and 
the grid spacing is well-matched to the final, binned resolution
of the WISP observations.

The origin of the coordinate system was
chosen following \cite{wes90} to lie at 
05$\mathrm ^h$ 24$\mathrm ^m$, $-$69$\arcdeg$
 50$\arcmin$ (B1950.0); this corresponds to 
the centroid of optical light in the galaxy (\cite{dev73}).  The distribution of
OB associations in the LMC, as in the Galaxy, can be assigned some
finite scale height above and below the galactic midplane.  However,
because this scale height is likely to be smaller than the scale
height of dust (\cite{har97}), and we have no {\it a priori} 
knowledge of the relative positions of each association along the
line of sight, we have forced the illuminating sources in our models
to lie in the plane of the LMC disk.  OB associations are not point
sources, having radii of $\approx$ 15--150 pc (\cite{luc70});
for simplicity, we have modelled them as spheres.  The radii of
our illuminators do not directly correspond to the optically
defined dimensions of the Lucke \& Hodge associations, but were 
determined from the vacuum ultraviolet images of SCH.

SCH photometered the entire Lucke \& Hodge 
catalog at 1500 and 1900 \AA, using rectangular apertures that
were matched to each association by hand; their Table 1 gives
the total area of each of their apertures.  We assigned radii
to our spherical sources by setting their projected surface
areas equal to the areas given by SCH in their Table 1.  In just two
cases (LH 15 and LH 77) did we deem the deviations from sphericality
strong enough to warrant a more complex procedure.  Both associations
lie within supergiant shells on the northern side of the LMC disk.
LH 15, within LMC-1, is contained within the field of view of the
WISP CCD image; LH 77, at the center of LMC-4, is quite bright and
resembles a quadrant of a circle's circumference.   These two 
associations were broken up arbitrarily into four identical
sub-associations, which more closely reproduced the visual appearance
of these sources.

Near-UV luminosities were assigned to the sources based on the 
photometry of SCH at 1500 \AA$ $ ($m_{15}$) and 1900 \AA$ $ 
($m_{19}$).  We dereddeded the SCH photometry and applied a correction
for the difference in bandpass between their filters and the WISP
filter at 2150 \AA.  The reddening values given by \cite{luc74}
were broken down into LMC and foreground Galactic components;
following SCH, the maximum value of foreground reddening was taken
to be E$_{B-V}^{MW}$ $=$ 0.07 mag.  Any additional reddening towards
the individual associations was attributed to dust within the LMC.
Reddening values for each source were derived following the procedure
outlined in Paper I, as were corrections for the differing bandpasses
used by SCH and in Paper I.
%
%
%
%
%
%

The derived extinction values were found to be in good
agreement with those published by \cite{smi90} in an erratum to SCH.
The bandpass corrections ranged from
$-$0.5 mag to $+$0.6 mag for the 122 Lucke-Hodge OB associations.
The corrected magnitudes were converted into monochromatic fluxes
for the Monte Carlo photon generator using the standard relation
F$_{UV} = 10^{-0.4(m_0 + 21.1)}$ (SCH).  The source positions,
radii, and luminosities are given in an appendix to this paper,
in Table A1.

\subsection{Dust Distribution}

The LMC is a disk galaxy and we have chosen to represent its
dust density distribution using an exponential decay with radius
and a hyperbolic secant law in height above the midplane (\cite{bin87}):

\begin{equation}
\rho(r,z) = A\; \exp(\frac{-r}{r_d}) \: \mathrm{sech}^2 
\mathnormal(\frac{z}{2z_d}),
\end{equation}

\noindent where the constant $A$ is set by the optical depth of the model
(see \S3 below), and the dust scale length $r_d$ and scale
height $z_d$ are taken from the literature.  Observational estimates
of the LMC's dust scale length were unavailable, and so we set 
$r_d$ $\equiv$ 2.6 kpc (12 grid units),
the scale length of the old stellar population
 (\cite{kin91})\footnote{We 
have assumed a distance to the LMC of 50 kpc}.

The dust scale height, $z_d$, must also be estimated indirectly.  
\cite{har97} measured two reddening-free photometric indices for
2069 O and B stars in a 2.9 deg$^2$ area centered
$\sim$ 2$\fdg$6 northwest of the optical center of the LMC.
Using the distribution of reddening values they found that the
data could be well-matched by a vertical distribution in which
the dust has a scale height equal to twice that of the OB stars.
Assuming the OB stars to lie in an extremely flattened disk,
with scale height $\approx$ 100 pc (c.f., \cite{oes95} for Galactic
OB stars), we choose a dust scale height of $\approx$ 200 pc, or
0.96 grid units.

%

\begin{figure*}
\centerline{\hbox{\psfig{figure=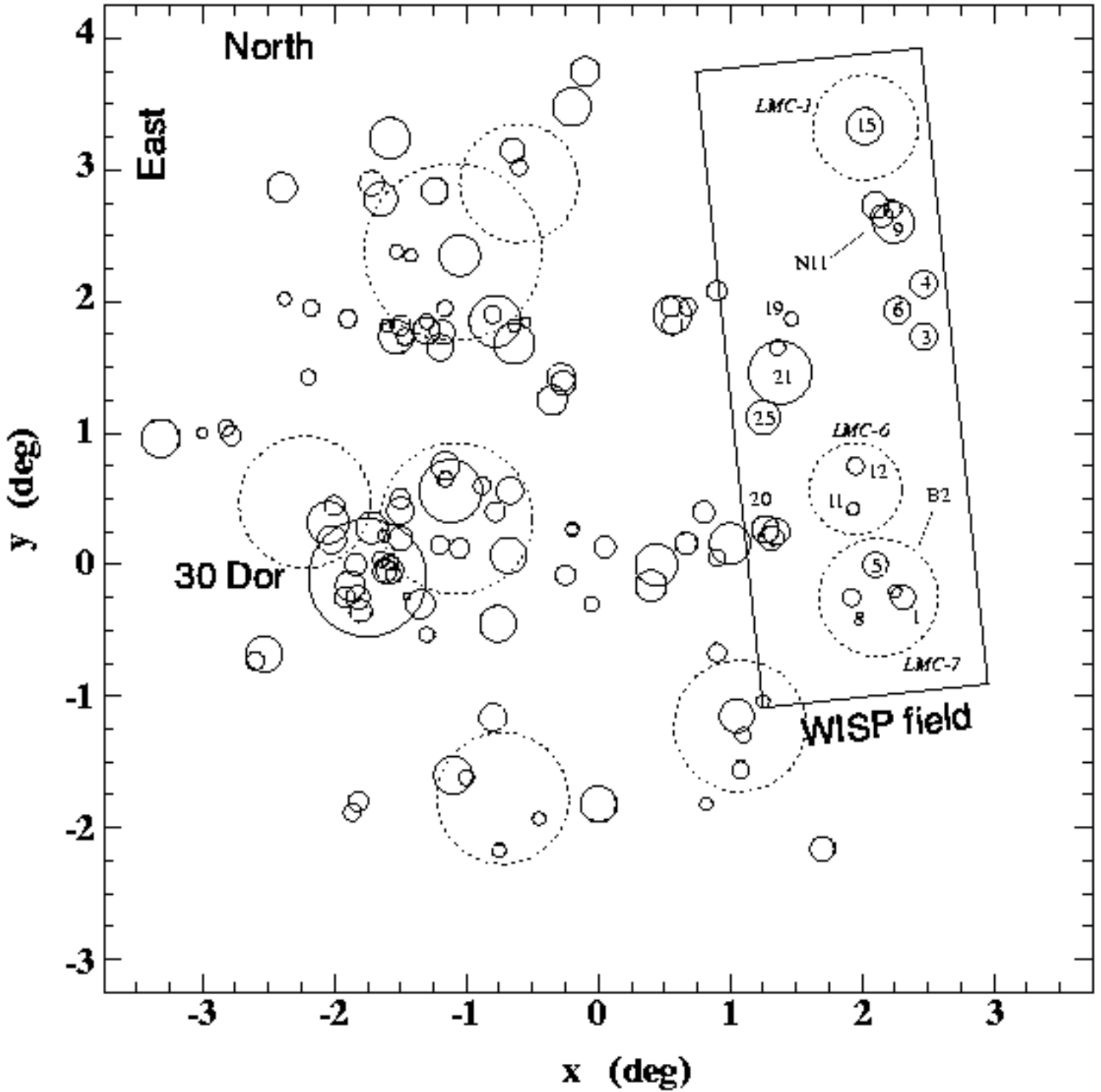}}}
\caption{The distribution of illuminators (open
circles) and cavities (dotted-line circles) in our model LMC.
The positions of illuminators are taken from \cite{luc70},
with radii and luminosities as described in \S2.1.  Cavity
positions are taken from \cite{mea80}.  Positions in equatorial
coordinates have been transformed onto a rectilinear grid with
the origin at
5$\mathrm ^h$ 24$\mathrm ^m$, $-$69$\arcdeg$ 50$\arcmin$\
(B1950.0).  Offsets are given in degrees.  North is up,
East is to the left.  OB Associations and supergiant shells
in the WISP field (Paper I) are labelled with their
Lucke-Hodge and Meaburn numbers, respectively.
\label{sources}}
\end{figure*}

As a first-order deviation from the smooth, azimuthally symmetric
model dusk disk, we placed nine low-density cavities into the model,
corresponding to the supergiant shells of \cite{mea80}.  These shells
were identified by the enhancements of H$\alpha$ emission around their
perimeters and are also characterized by extremely low H{\small I}
column densities.  They are thought to be roughly cylindrical,
and ``open-topped'' (\cite{wes97}), but
in our models they are defined by simple spherical 
cavities of low optical depth.  The cavities are placed in the 
midplane of our model galaxy, using the positions and sizes from
\cite{mea80}.  We assign a density to the cavities by defining the
near-UV optical depth $\tau_c$ across the diameter of a cavity.
$\tau_c$ was derived from photometric measures of the reddening,
E$_{B-V}$, of the OB associations lying within the boundaries of
the supergiant shells.  These lie in the range 0.00 $\leq$ 
E$_{B-V}$ $\lesssim$ 0.12 (\cite{luc74}), less 0.07 mag of 
foreground reddening; we also assumed that roughly half of the
observed reddening towards the OB associations was due to material
in the near neighborhood of the stars and hence not a contributor
to the optical depth of the cavity as a whole.   We adopted 
a ``typical'' E$_{B-V}$ of 0.01 mag, and assumed the OB associations
to lie at the center of the spherical cavities; applying an LMC
extinction law for the model's 2150 \AA$ $ photons, we set
$\tau_c$ = 0.1.  The catalog of supergiant shell parameters is
given in the Appendix, in Table A2; the cavities and illuminators
are mapped out in Figure \ref{sources}.

The inclination of the LMC disk to the 
plane of the sky remains a matter of some debate (\cite{wes97}).
It has become clear that the east (30 Dor) side of the LMC is 
closer than the western (WISP field) side.  As shown in Table 3.5
of \cite{wes97}, both the inclination i and position angle line
of nodes $\Theta$ are known only to a precision of a few tens
of degrees.  Measurements of $\Theta$ 
scatter around a north-south line, and so we adopt $\Theta = 
180\arcdeg$ for simplicity.  Geometrical methods applied to 
young and old stellar populations as well as neutral and ionized
gas have yielded results varying between 25$\arcdeg \lesssim$ i 
$\lesssim 48\arcdeg$.   The expected magnitude and spatial variation of
polarization depend strongly on the scattering geometry in the 
disk of the model galaxy, and so we consider three 
values for i in our models: 28$\arcdeg$, 36$\arcdeg$, and 45$\arcdeg$.
The WISP field, along the west side of the LMC, is tilted away
from the Earth: as the inclination increases,
photons must traverse larger path lengths through the absorbing
dust layer in order to escape and be seen.


\section{Dust Properties}

Along with the scattering geometry, the optical properties of
interstellar dust grains control the linear polarization of
stellar photons.  Our models treat the scattering process using
a standard Henyey-Greenstein phase function (\cite{hen41}), which
depends on the albedo $a$, and asymmetry parameter $g$.
The parameter $g$ defines the probability
for an incident photon to scatter through an angle $\theta$:
P($\theta$) $\propto$
$\frac{(1 - g^2)}{(1+g^2-2g\cos \theta)^{\slantfrac{3}{2}}}$.
$g$ = 0 yields isotropic scattering, while $g$ = 1 gives pure
forward scattering; $g$ $<$ 0 corresponds to backscattering.
To model the polarization, we follow White (1979) in the 
approximation P($\theta$) $\approx$ $p_{\mathrm{max}}$
$\frac{\sin ^2 \theta}{1 + cos ^2 \theta}$.  $p_{\mathrm{max}}$
is the maximum polarization attainable in a single scattering
event, for a scattering angle of 90$\arcdeg$; the polarization
of the scattered photon decreases for smaller and larger scattering
angles.

The total amount of dust in the model is
described by the single optical depth parameter $\tau_{eq}$.
$\tau_{eq}$ is simply the optical depth of the model galaxy to
a photon as it travels from center to edge through the midplane.

\cite{whi79} tabulated the scattering properties of the \cite{mat77}
(MRN) Milky Way dust mixture.  Deviations from the \cite{whi79} values
are to be expected for the LMC, which is in general more metal-poor
than the Galaxy (\cite{pag78}; \cite{duf84}).  The LMC extinction 
curve shows a less pronounced 2175 \AA$ $ bump and a steeper rise into
the far-UV than does the Galactic curve, attributable to variations
in dust grain sizes and/or compositions (\cite{nan81}).  \cite{pei92}
recalculated the albedo of Magellanic Cloud dust using an
MRN grain-size distribution with the relative contributions from
graphite and silicates scaled to match the observed mean extinction
curves.  We adopt the Pei (1992) value, $a$ = 0.66 for $\lambda$ = 2150 \AA, 
for all our models; this is $\approx$25\% higher than the observed albedo
of Milky Way dust (e.g., \cite{wit92}).  \cite{pei92} did not include
calculations for $g$ or $p_{max}$ in his paper; 
we adopt the MRN-based value, $p_{max}$ = 0.31.  Because the polarization
of scattered starlight
depends strongly on $g$, we compute families of models in which $g$ is
permitted to vary.

In the ultraviolet, the phase function asymmetry parameter $g$ is poorly constrained
by both models (0.1 $\lesssim$ $g$ $\lesssim$0.7; e.g., \cite{mur94}),
and observations, (0.3 $\lesssim$ $g$ $\lesssim$0.9; e.g., \cite{sas96}),
even for Milky Way dust.
Variations in $g$ lead to differences in the 
expected polarization patterns, allowing us to infer its value through
comparisons of the model output to observations.   We chose to examine
the moderately to strongly forward-throwing regime: $g$ = 0.64, corresponding
to the MRN value, and $g$ = 0.90, suggested by observations of
reflection nebulae in the Pleaides cluster (\cite{gib97}).  Intermediate values were
chosen at $g$ = 0.70, 0.77, 0.83, close to the values derived by
\cite{wit92} for the reflection nebula NGC 7023.  The most recent models of the
Pleiades nebulosity (Gibson \& Nordsieck 1999, in preparation) also
indicate a moderate value for $g$. 


For the LMC, the parameterization of dust mass using 
$\tau_{eq}$ is problematic, because
the LMC is observed nearly face on and thus the derived value of 
$\tau_{eq}$ is strongly geometry-dependent.  We derive an initial optical
depth from observations of OB associations (\cite{har97}), scaled by
a geometric factor, and then correct this value using the observed
ultraviolet (Paper I) to far-infrared (\cite{deg92}) flux ratios.
\cite{har97} found a mean
B-band optical depth through the face-on LMC $\tau_B$ = 0.98 $\times$ cos i.
Using the LMC mean extinction curve of \cite{fit86}, the 2150 \AA$ $ optical
depth to the LMC's midplane is $\tau_{UV}$ = $\frac{1}{2} \times
2.4 \tau_B$ = 1.2 $\times$ cos i.  The \cite{har97} field is offset
some 0.67 radial scale lengths from the centroid of our dust distribution,
and so we adopt a pole-to-midplane central optical depth $\tau_{pole}$ = 
2.4 $\times$ cos i.  $\tau_{eq}$ is finally obtained by scaling $\tau_{pole}$
by the flattening ratio of the disk.  We adopt as our initial value:

\begin{eqnarray}
\tau_{eq}\; \equiv\; \tau_{pole} \frac{r_d}{z_d}\; =\; 29.5 \cos i.
\end{eqnarray}

This number was adjusted during the modelling procedure in 
order to match the observed ratio of WISP 2150 \AA$ $ flux to
IRAS 60 $\mu$m flux (see \S5, below).

\section{Radiation Transfer}

We construct model scattered light images with a Monte Carlo continuum
radiation transfer code which accounts for multiple photon scattering,
and predicts the spatially resolved flux and polarization (\cite{woo97}).
In the radiation transfer calculation,
the dust plus gas mixture has albedo, $a$, and a scattering
phase function approximated by the Henyey-Greenstein phase function
(\cite{hen41}) with asymmetry parameter, $g$ (section~3).
The code has been modified from the axisymmetric
models of Wood \& Jones to include a three dimensional distribution of
illuminating sources (section~2.1) --- this is crucial for modeling the
UV scattered light pattern in the LMC.
Additionally, we have added a new feature, where absorbed photons
are not removed from the simulation, but ``re-radiated'' isotropically
from the point of absorption to form a ``far-infrared'' image.  
This is a first approximation for
predicting the far IR emission from our simulations.  We are in effect
assuming that all the UV emission that is absorbed is reradiated at one
wavelength where the dust is optically thin.  In order to compare our models
to the WISP and IRAS images of the LMC, we must include additional factors
that account for the fact that our ``reprocessing '' technique does not
enforce radiative equilibrium (section~5).

\section{Modelling Procedure}

With the dust distribution and OB association properties held 
constant for all models, our goal is to find the combination 
of disk inclination and scattering phase function that best
reproduces the observed polarization maps of Paper I.
The polarimetric images of the LMC are the result of the
interplay of a large number of physical processes and properties:
the vertical, radial, and azimuthal dust distributions, the 
clumpiness of the dust, grain albedos, the distances above or
below the midplane of each illuminating source, and the 
morphologies and densities of the supergiant shells all contribute
to the observed polarization pattern. 

By tuning all of these
quantities independently in our models, we {\it could} create
a model whose output reproduces the observations precisely.
However, such a model would almost certainly be non-unique, and
its physical reality would be highly questionable.  We prefer
instead to hold most of the model traits fixed and consider
only the variation in the parameters that are expected to 
most strongly influence the polarization maps.  For example,
one of the observables we insist the models match is the
ratio of ultraviolet to far-infrared light escaping from the
galaxy.  In reality, this obviously depends on the UV albedo $a$,
but because $\tau _{\mathrm{IR}}$ $\ll$ 1 and $\tau _{\mathrm{UV}}$
$>$ 1, it also depends on the optical depth.
We could trade off albedo and optical depth to tune
this ratio precisely, but the model would suffer from severe
non-uniqueness and could be driven to unphysical values of
$a$, $\tau$, or both.  Therefore we fix $a$ (see Section 3),
and adjust $\tau$ until the model produces approximately the
correct amount of infrared light.

We consider a grid of 15 models (see Table 1), comprising
3 values of disk inclination i, and 5 values for scattering phase
function $g$.  For each model, we perform an initial run in 
order to determine the correction to our first guess at $\tau_{eq}$.
If the emitting dust is smoothly distributed
and in radiative equilibrium, then the relation between absorbed
starlight and thermal emission in the far-infrared depends simply
on the dust optical depth.  Hence we can attempt to match the 
amount of dust in our models to the true dust mass by a comparison
of UV to IR fluxes; we correct our intial value of $\tau_{eq}$
by matching the model images in the UV and IR to the observed
WISP 2150\AA$ $ and IRAS 60 $\mu$m images.  

\begin{deluxetable}{ccccc}
\tablewidth{0pt}
\tablenum{1}
\tablecaption{Assumed, derived, and output parameters for
our Monte Carlo models.}
\tablehead{
\colhead{$g$\tablenotemark{a}} &
\colhead{i\tablenotemark{b} $ $ (deg)} &
\colhead{$\tau_{eq}$\tablenotemark{c}} &
\colhead{$\langle p \rangle$\tablenotemark{d} $ $ (\%)} &
\colhead{$\delta$x\tablenotemark{e} $ $ (arcmin)}}
\startdata
0.64 & 28 & 24.8 & 12.2 $\pm$0.3 & 0.6 $\pm$0.6 \nl
0.64 & 36 & 22.4 & 12.4 $\pm$0.3 & 10.2 $\pm$0.6 \nl
0.64 & 45 & 20.1 & 12.8 $\pm$0.3 & 16.2 $\pm$0.6 \nl
0.70 & 28 & 25.9 & 12.7 $\pm$0.3 & -0.6 $\pm$0.6 \nl
0.70 & 36 & 23.3 & 12.6 $\pm$0.3 & 9.6 $\pm$0.6 \nl
0.70 & 45 & 19.9 & 12.9 $\pm$0.3 & 15.6 $\pm$0.6 \nl
0.77 & 28 & 25.5 & 12.9 $\pm$0.3 & 0.0 $\pm$0.6 \nl
0.77 & 36 & 23.5 & 12.8 $\pm$0.3 & 8.4 $\pm$0.6 \nl
0.77 & 45 & 19.8 & 13.0 $\pm$0.3 & 16.2 $\pm$0.6 \nl
0.83 & 28 & 25.3 & 13.1 $\pm$0.3 & -0.6 $\pm$0.6 \nl
0.83 & 36 & 24.5 & 13.4 $\pm$0.3 & 6.6 $\pm$0.6 \nl
0.83 & 45 & 20.3 & 13.2 $\pm$0.3 & 15.6 $\pm$0.6 \nl
0.90 & 28 & 26.5 & 13.4 $\pm$0.3 & -1.2 $\pm$0.6 \nl
0.90 & 36 & 24.4 & 13.7 $\pm$0.3 & 7.2 $\pm$0.6 \nl
0.90 & 45 & 20.8 & 14.1 $\pm$0.3 & 13.8 $\pm$0.6 \nl
\hline
\multicolumn{3}{l} {\bf Observations\tablenotemark{f}} &
12.6 $\pm$2.3 & 8.4$^{+1.8}_{-5.4}$ \nl
\enddata
\tablenotetext{a}{Scattering asymmetry parameter, see \S3.}
\tablenotetext{b}{Inclination angle of LMC disk, see \S2.2.}
\tablenotetext{c}{central optical depth in the plane.}
\tablenotetext{d}{mean percentage of linear polarization.}
\tablenotetext{e}{offset in polarization centers of
symmetry, west of illuminating source.}
\tablenotetext{f}{from Cole {\it et al.} 1999a.}
\end{deluxetable}

Rather than resort to large-scale averaging over extended regions
of the LMC, we choose instead to calibrate our model optical
depths using one well-measured region that lies within the
WISP field of view (Paper I): the N11 complex (also
known as DEM 34 [\cite{dav76}], B1 [\cite{mar76}], MC18 
[\cite{mcg72}]).
N11 lies at the southern edge of the supergiant shell LMC-1
(\cite{mea80}), and
contains a large H {\small II} region that is ionized by hot 
stars in OB associations LH9, LH10, LH13, \& LH14 (\cite{luc70}).
This permits us to make accurate comparisons between our models
and polarimetric observations.

The correction is complicated because of the mismatch between
models and reality.  From the models, we compare monochromatic
stellar photons at 2150 \AA$ $ to a monochromatic far-infrared
emission from dust which is heated by the starlight.
From observations, we compare images in the $\lambda$ $\approx$
2150 \AA$ $ bandpass to emission at 60 $\mu$m; real dust is heated
by starlight of all wavelengths.  Therefore the model IR/UV ratio
cannot be immediately compared to the observations.  In order
to find the appropriate dust optical depth, we require:

\begin{eqnarray}
\frac{\mathcal{F}\mathrm{_{IRAS}}}{\mathcal{F}\mathrm{_{WISP}}}\;
=\; \varepsilon^{UV}_{IR}\; \eta^{UV}\; \eta^{IR}\; 
\frac{\mathcal{F}\mathnormal{_{IR}}}{\mathcal{F}\mathnormal{_{UV}}}, 
\label{calib}
\end{eqnarray}

\noindent where $\mathcal{F}\mathrm{_{IRAS}}$ and 
$\mathcal{F}\mathrm{_{WISP}}$ are the observed fluxes in the 
IRAS 60 $\mu$m and WISP 2150 \AA$ $ bandpasses, and 
$\mathcal{F}\mathnormal{_{IR}}$ and $\mathcal{F}\mathnormal{_{UV}}$ are the 
fluxes in the model output IR and UV images.  

$\eta^{IR}$ is a correction factor to account for the fact that
our model dust grains are not in thermal equilibrium; they
radiate their absorbed energy at a single, average, far-infrared
wavelength whose flux equals the bolometric far-IR flux
of the dust.  $\eta^{IR}$ depends on the dust temperature and
the wavelength dependence of the dust emissivity; we adopt 
the values from \cite{deg92} in her calculation of the ionizing
flux in N11.

$\eta^{UV}$ is a similar correction that relates the dust heating
by photons in the WISP bandpass to the total dust heating from
light of all wavelengths.  We calculate $\eta^{UV}$ by integrating
the light of N11 from 912 \AA$ $ to 3648 \AA$ $ using Kurucz model
atmospheres, measurements of the initial mass function from \cite{deg92}
(also see \cite{par98}), and weighting the spectral energy distribution
by an LMC extinction law.

$\varepsilon^{UV}_{IR}$ corrects for the fact that our dust grains
are not in radiative equilibrium; the model re-emits one far-infrared
photon for each absorbed near-ultraviolet photon; to conserve energy
we must scale the model output images by the ratio of UV to IR mean
photon energies, represented by $\varepsilon^{UV}_{IR}$.

To tune the optical depth of our models, we checked each output
model against equation \ref{calib}; where the model ratio exceeded
the the observations, we lowered $\tau_{eq}$, and vice versa.
By lowering the UV optical depth, we decreased the number of
reflected photons that are subsequently 
absorbed, and hence the relative amount of far infrared emission.
As a result of the tests, we adopted the parameters shown in 
Table 1 for each inclination; no systematic trend with
$g$ was apparent, although the values of 
$\frac{\mathcal{F}\mathnormal{_{IR}}}{\mathcal{F}\mathnormal{_{UV}}}$
showed a scatter of $\pm$10\% around the mean value at each inclination.
Due to difficulties with the IRAS zeropoint calibration, and 
uncertainty in the WISP zeropoint, it is difficult to relate our
model parameter
$\frac{\mathcal{F}\mathnormal{_{IR}}}{\mathcal{F}\mathnormal{_{UV}}}$
to a physical flux ratio.




Because of the large number of photon sources, the number of output
images, and the large number of pixels in each output image, large numbers
of photons were required in order to obtain significant signal-to-noise
to measure the polarization of the diffuse UV light.  For each of the
15 permutations of i and $g$, we computed a ``low'' signal-to-noise
model with 10$^8$ photons propagating through the model galaxy.
These models provided sufficient information to identify the models
that matched the data well enough to merit a more detailed look.

We re-ran the best fitting model with 10$^9$ photons in order
to more accurately assess the mean level of polarization and track
the variation in polarization level and position angle across the 
image.  As we began to write up these results, we continued to let
the model run in order to create the highest possible signal-to-noise
in the output images.

\section{Results \& Discussion}

\subsection{Model Images \& Polarization Maps}

Our models produced as their output a set of four images for each
run:  the ultraviolet and infrared flux, and, in the ultraviolet,
images of the linearly polarized flux Q and U.   For analysis, the
output data were converted into FITS format and examined within 
the IRAF\footnote{IRAF is distributed by the National Optical
Astronomical Observatories, which are operated by the 
Association of Universities for Research in Astronomy, Inc., 
under cooperative agreement with the National Science Foundation.} suite of tasks.  

\begin{figure*}
\centerline{\hbox{\psfig{figure=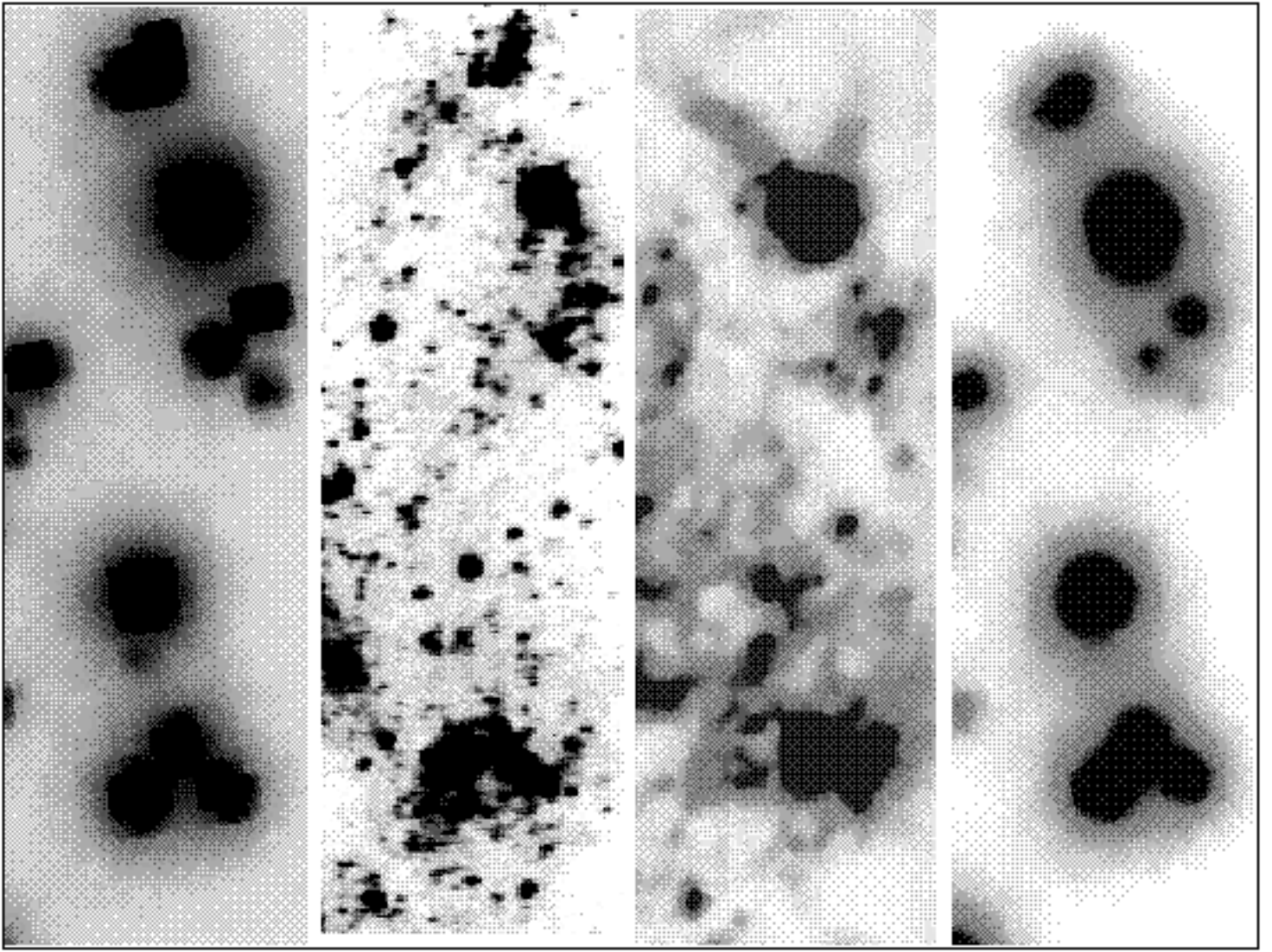}}}
\caption{Model output images for i = 36$\arcdeg$,
$g$ = 0.70, and 4.1 $\times$ 10$^9$ total photons.  From left to
right: model UV image; WISP 2150 \AA$ $ image (Paper I); 
IRAS 60 $\mu$m image (\cite{sch89}); model IR image.
North is up, East is to the left.  The differences between model
and observations are attributable to the highly non-uniform distribution
of stars and dust in the LMC.
\label{quad}}
\end{figure*}

Model images of the WISP-observed portion of the LMC are shown in 
Figure \ref{quad}.  From left to right, we show the model UV image,
the WISP 2150 \AA\ image, the IRAS 60 $\mu$m image, and the model
IR image.
The far left and right panels of Figure \ref{quad}
show the co-added results of multiple Monte Carlo runs, amounting to a
total photon count of 4.1 $\times$ 10$^9$.  Only a portion of the image is
shown, to facilitate comparison to the WISP data (middle left panel of
Figure \ref{quad}).
The dynamic range of the UV model is significantly 
smaller than the 2150 \AA\ data, because of the uniformly high
optical depth in our models which results in nearly constant
attenuation across the field.
Many of the features are well-reproduced by 
the models, including the prominent N 11 and B2 complexes which
particularly dominate the IRAS 60 $\mu$m image.
The supergiant shells are less visible than expected in the model
IR image; it lacks the ``holes'' visible in the 60 $\mu$m data.
However, their effect can be noted
by comparison of the UV and IR appearance of the LH15 association
at the upper right of the panels in Figure \ref{quad}: LH15 is far brighter
in the UV than the IR, a consequence of its location within the 
shell LMC-1 (see Figure \ref{sources}). 

Despite the obvious, and expected,
lack of fine-scale structure in our model, we nonetheless are 
able to approximate the large-scale luminosity distribution in 
these two wavelength regimes.  The most obvious failure of our
model to reproduce small-scale structure in the dust distribution
occurs at the association LH12 (see Figure \ref{sources}). 
LH12 is an intrinsically
bright association (SCH) that is highly reddened (Lucke 1974) and
so appears faint to observers.  However, LH12 lies within the LMC-6
supergiant shell; in our simple model, this drastically reduces 
the amount of obscuring dust along the LH12 sightline, causing us
to dramatically overestimate the observed brightness of the association.
A similar effect is discernible for the more southerly association
LH5, highly reddened despite its position near the edge of
the LMC-7 supergiant shell.

We created images of the degree of
linear polarization, $p$, and position angle $\theta$:

\begin{eqnarray}
\mathrm P\; =\; \left( \frac{(Q^2\; +\; U^2)}{I^2} \right) ^{0.5}\\
\theta \; =\; 0.5\; \arctan \left( \frac{U}{Q} \right) .
\end{eqnarray}

\begin{figure*}
\centerline{\hbox{\psfig{figure=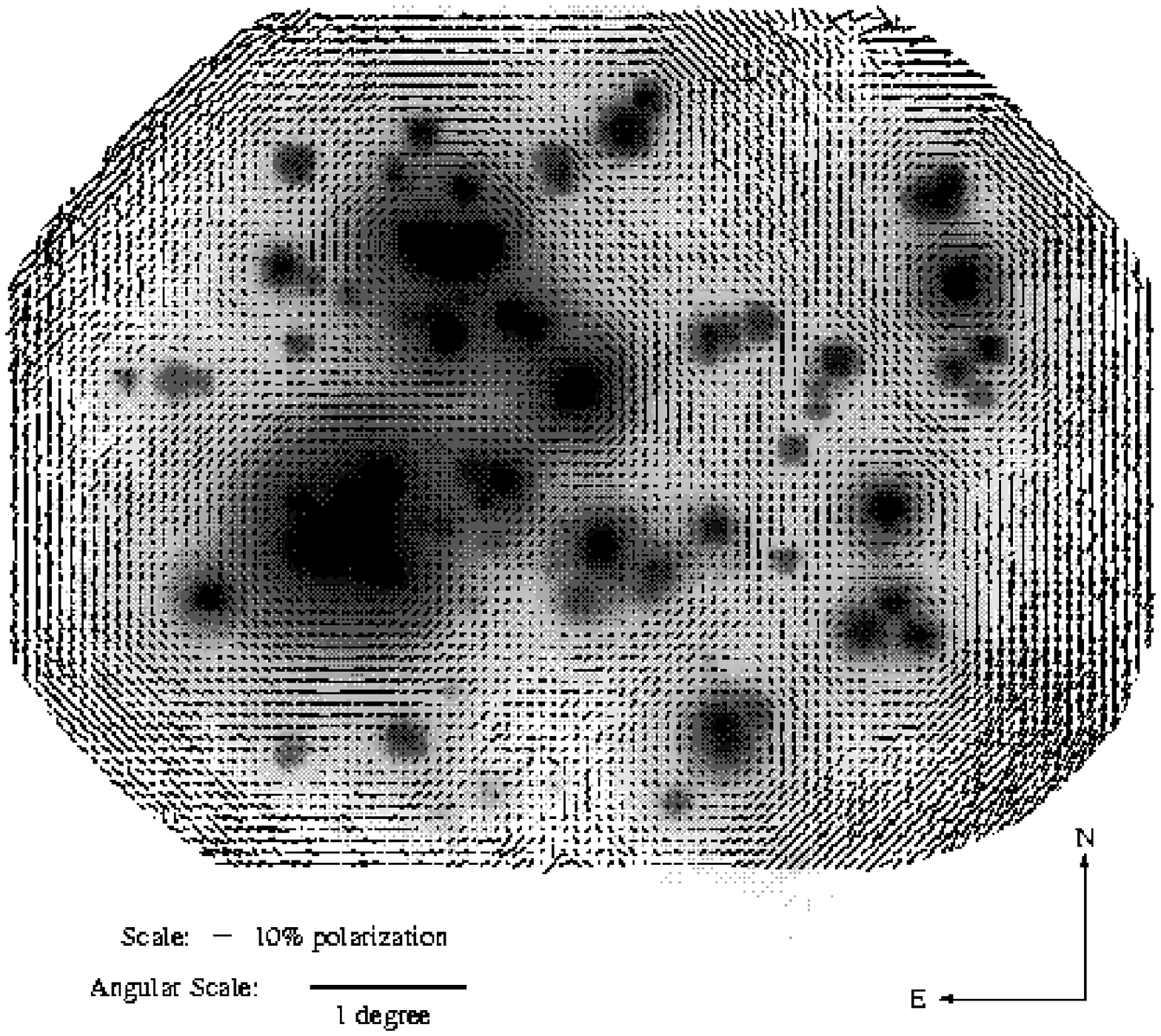}}}
\caption{Model UV image with the 
polarization vectors derived from the
Stokes Q and U model images overplotted.  Because the only illumination
derives from the OB associations, and the dusty scattering medium is
smoothly distributed, the polarization vectors present clear, regular
centrosymmetric patterns about the illuminators.  North is up, east is
to the left.
\label{model}}
\end{figure*}

We created polarization maps
with vector length proportional to P and position angle equal
to $\theta$; our highest S/N map is shown in Figure \ref{model}.  The 
smooth dust distribution and finite number of illuminating sources
account for the extreme regularity of the model polarization map.

\begin{figure*}
\centerline{\hbox{\psfig{figure=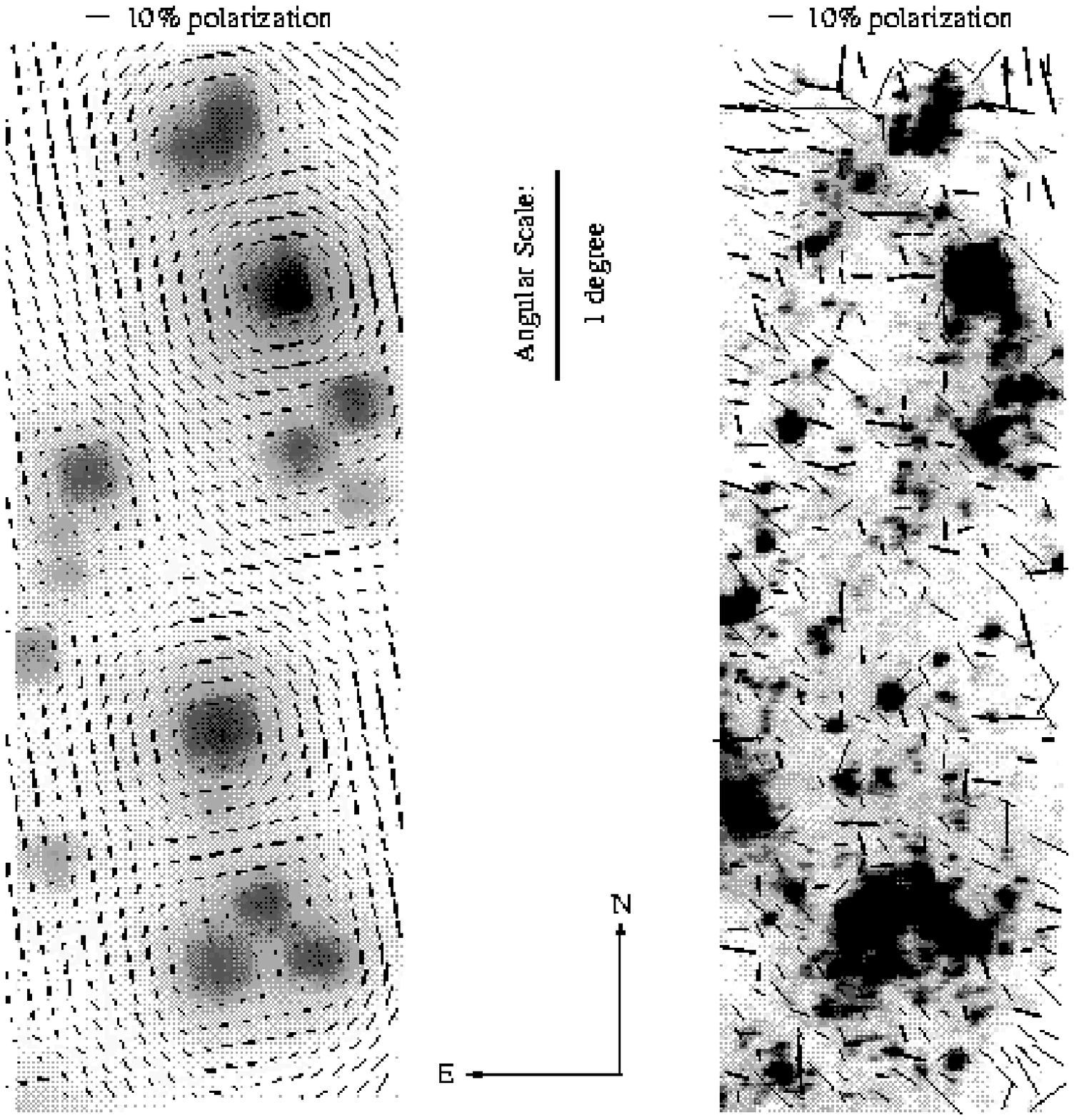}}}
\caption{{\sl left:} model UV image with
polarization vectors overplotted;
{\sl right:} WISP 2150 \AA$ $ image with polarization vectors.
The polarization and angular scales are shown in the figure.
\label{wisp}}
\end{figure*}

In order to constrain the disk
inclination of the LMC and the phase function $g$ of its dust,
we compared the model polarization maps to those produced by
Paper I from the WISP observations, shown side-by-side in
Figure \ref{wisp}.  
The WISP observations, binned into 6$\arcmin$ $ $ pixels to
increase signal-to-noise, showed some evidence for the presence
of centro-symmetry about the brightest OB associations, but 
were hampered by the shortage of photons far from these bright
regions; the effects of inhomogeneities in the scattering
medium and a diffuse starlight component clearly dominate
the appearance of the observed polarization maps. 

Our points of comparison included the mean P and $\theta$, 
the shape of the high-polarization tail of P, the degree of
centro-symmetry around the brighest regions (N11 [B1] in North
and N79 [B2] in the South), and the offsets of the polarization
symmetry centers from the central OB associations.
We also looked for variations in mean P and $\theta$ across
the $1\fdg5$ $\times$ $4\fdg8$ field of view, but found no
statistically significant differences to distinguish the various
models.
A

The comparison was not straightforward because of the systematic
effects which dominate the errors in the observed polarization
maps.  In particular, the mean value of P was determined quite
carefully: the WISP maps have been cleaned of marginal polarization
detections where P $<$ 2.5$\sigma _{\mathrm P}$, and become incomplete for
polarizations less than 10\%.  Therefore we applied a linear
incompletess correction to the models, such that the model
polarizations are weighted progressively less between 10\% and
4\%, and all polarizations smaller than 4\% are ignored.  This
mimics the observational bias towards detection of regions of
high polarization.

The observed polarization maps, biased by incompleteness toward
the detection of high polarization regions, showed a much higher
scatter in polarization values than did the models.  The observations
also show a patchier distribution of P; pixels with polarizations
in the 5--10\% range are frequently juxtaposed with $\approx$20\% 
polarized regions.  This indicates
that the effects of small, optically thick clumps in the ISM are strongly
influencing the scattered light component of the diffuse UV background.
However, when we binned the models to the resolution of the observations
and corrected for incompleteness at P $<$ 10\%, our models were 
consistent with the observed mean level of polarization across the WISP
field.  Scattered light from OB associations is indeed likely to 
account for a large fraction of the diffuse ultraviolet background
in the LMC.

\subsection{Disk Inclination and the Dust Scattering Phase Function}

\begin{figure*}
\centerline{\hbox{\psfig{figure=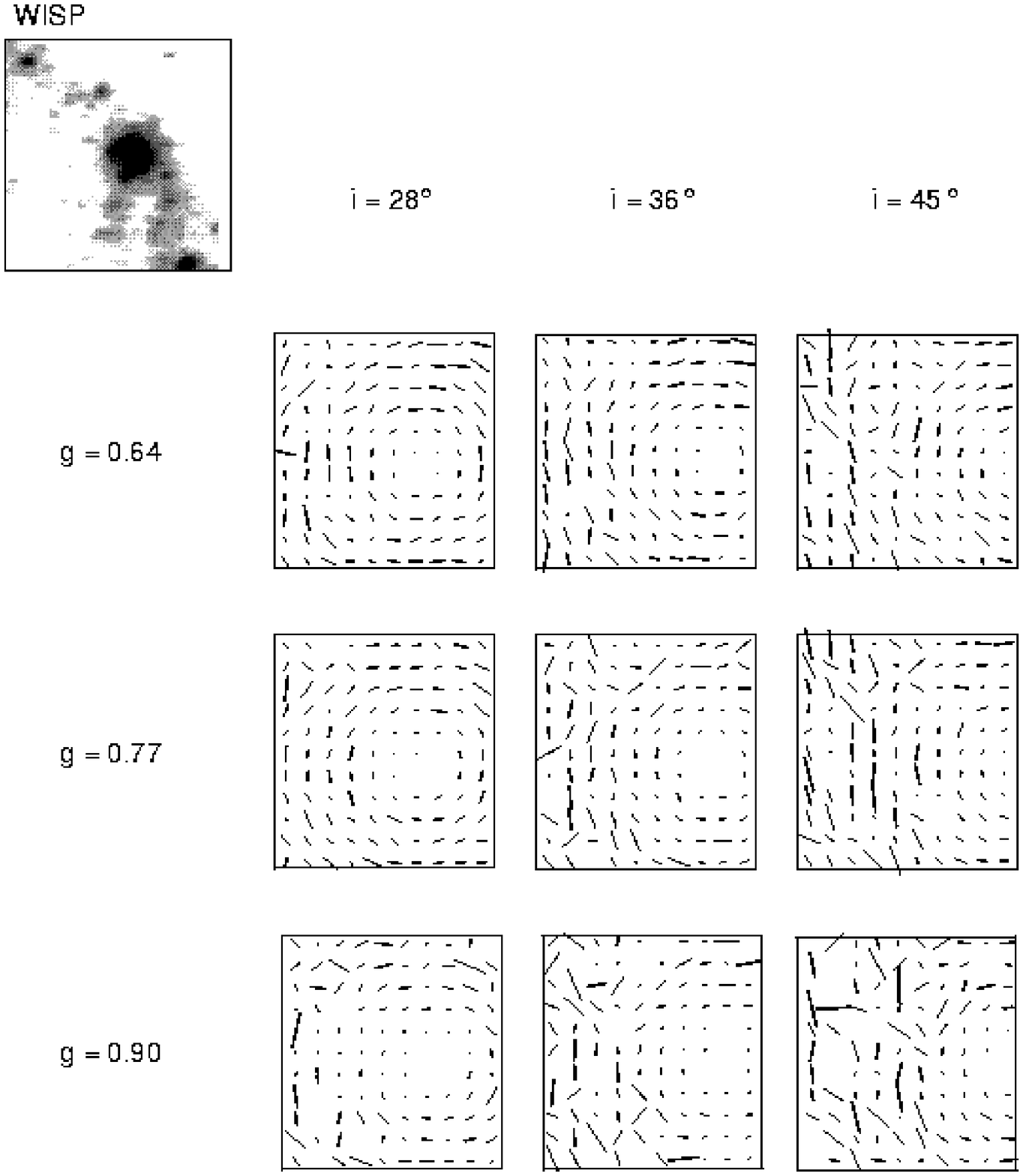}}}
\caption{Model polarization maps for the area surrounding the
H {\small II} complex N 11 demonstrate the dependence on inclination
and phase function asymmetry.  The actual 2150
\AA\ image of the region, 1$\arcdeg$
on each side, is shown at upper left.  Polarization maps for the same 
area are shown, arrayed by inclination and $g$ value.  The smallest 
polarization vectors plotted are 1\%, the largest 35\%; observed values
around N 11 range from 0 to $\approx$10\%.
\label{n11}}
\end{figure*}

\begin{figure*}
\centerline{\hbox{\psfig{figure=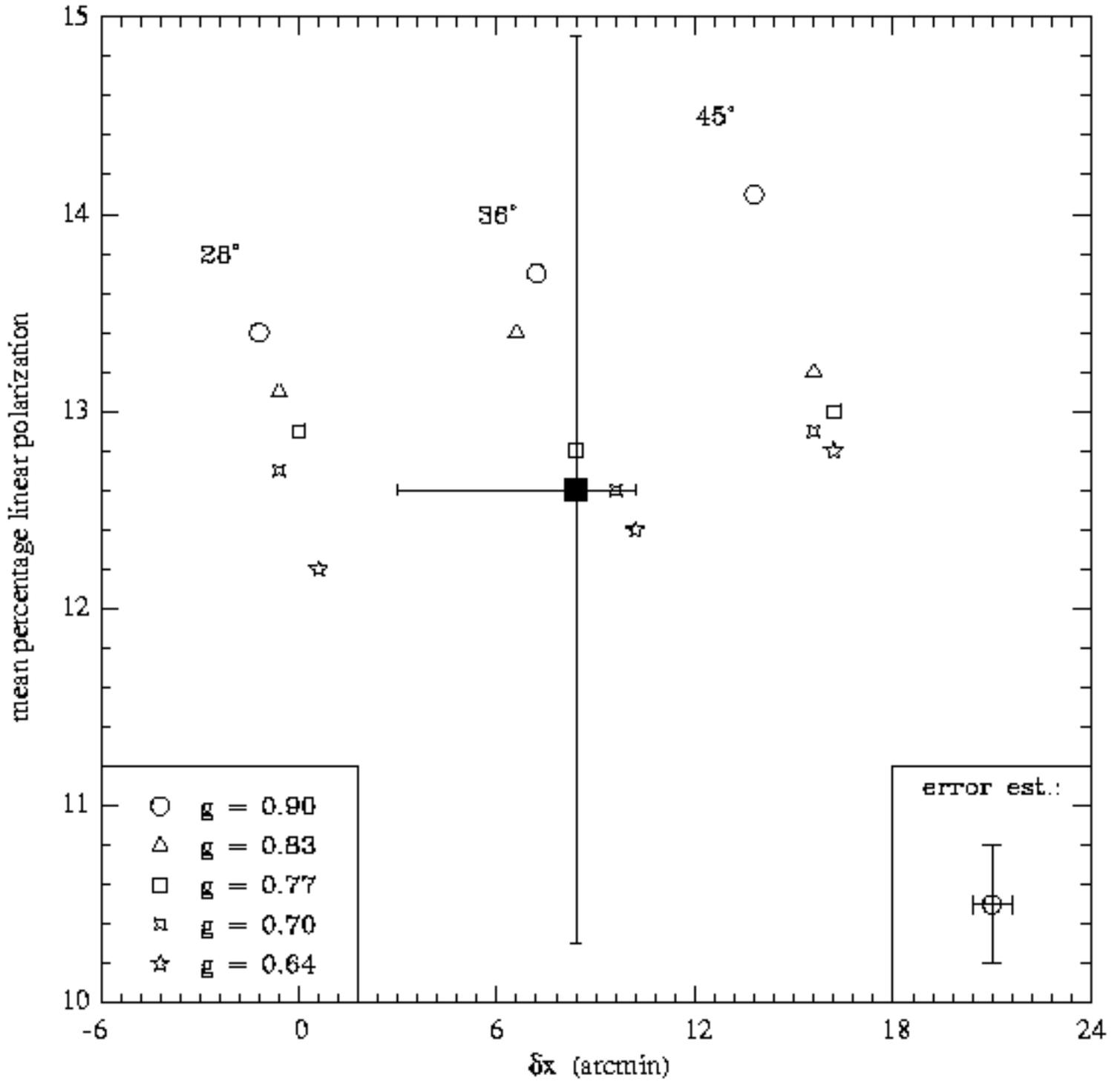}}}
\caption{Model images created with differing values
of i and $g$ separate themselves in a plot of mean percentage of 
polarized flux against the offset between the observed scattering
halos and illuminating sources.  $\delta$x is shown in arcminutes.
The solid square shows the
observational data with errorbars (Paper I).  A typical errorbar
for the model-derived points are shown in the lower right corner.
Each nearly vertical column of symbols is labelled with the disk
inclination; the legend relating symbol type to $g$ is at lower left.
\label{result}}
\end{figure*}

We found that our model polarization maps were sensitive to variations
in $g$ and i; see Figure \ref{n11}.  As $g$ increased from 0.64 to 0.90,
the mean polarization $\langle \mathrm P \rangle$ 
increased from $\approx$12.5\% to $\approx$13.7\%.  The
disk inclination manifested itself most noticeably in the distribution
of polarization vectors around bright sources, e.g., N11.  For a 
face-on disk of scattering material, the polarization vectors form 
a centro-symmetric pattern about the illuminating source; the inclination
introduces an asymmetry which shifts the center of the distribution away 
from the illuminator, perpendicular to the disk's line of nodes.  As
the inclination was increased, the offset in symmetry
center of the polarization patterns (hereafter referred to as 
$\delta$x) varied from 0$\farcm$36 eastward, at 28$\arcdeg$, 
to 16$\farcm$2 westward, at 45$\arcdeg$.
These results are summarized in 
Table 1, and plotted in Figure \ref{result},
in which $\langle \mathrm P \rangle$
and $\delta$x are plotted for each
model.  We have also plotted $\langle \mathrm P \rangle$ 
and $\delta$x for the observed
polarization map, with the associated error bars.  $\delta$x
tends to zero at inclinations of $\lesssim$30$\arcdeg$ due
to the competing effects of the disk inclination and the
radial drop in dust density from east to west across the 
field; a plane parallel slab of scattering dust would
show $\delta$x = 0 only for i = 0.

Interpolating in inclination, and adopting the uncertainty in our
observational determination of $\delta$x from the binned WISP data,
we find that the LMC's disk is inclined at 36$\arcdeg^{+2}_{-5}$
to the plane of the sky.  Insofar as the western side of the LMC
resembles our models, this is a direct, geometric determination of
its inclination.  Although inconsistent with recent kinematic
determinations (e.g., \cite{kim98}, who found i = 22$\arcdeg$ 
$\pm$6$\arcdeg$), it is squarely in agreement with the average
of reliable values tabulated by \cite{wes97}.  If the dust scale
height is larger than we have assumed, then the inclination required
to yield a given $\delta$x is reduced; a significant change in 
$z_d$ would be required due to the concentration of scattering
dust close to the midplane of the disk.

It is more difficult to draw conclusions regarding $g$.  As
seen clearly in Figure \ref{result}, the effect of high $g$ is mainly
to increase the mean level of polarization.  This ran counter
to our expectations, for a high value of $g$ should produce
smaller scattering angles, on average, resulting in polarizations
much smaller than $p_{max}$.  The counterintuitive result is
due to our simulation of the observational biases in the data
of Paper I (see section 6.1); the high $g$ models did initially
show lower levels of polarization, but were cleaned more severely
by our rejection of low P pixels.  

The low signal-to-noise of the WISP data 
and the larger than expected polarization fluctuations
result in 
uncomfortably large errors in our determination of
$\langle \mathrm P \rangle$.  At the 3$\sigma$ level,
we cannot rule out any value of $g$ between 0 and 1.  At the 1$\sigma$
level, we find $g$ = 0.7 $\pm$0.3.  The simple comparison of
$\langle \mathrm P \rangle _{mod}$ to $\langle \mathrm P \rangle _{obs}$
throws away information about the distribution of
$\langle \mathrm P \rangle$ within the
images: we found that for $g$ = 0.83--0.90, our models
exhibited a strong high-P tail extending to $\approx$30\%.  This tail
was weak in the observations, and suggests that such high values
of $g$ are less likely than would be inferred from the
formal error distribution.  However, as noted in Paper I, the 
highest polarization pixels typically show low flux values, and
hence relatively large errors in both P and $\theta$.  Low 
signal-to-noise might therefore introduce a bias against the 
high-$g$ models.  

Our determination of $g$ is of comparable
precision to values for the diffuse Milky Way dust, e.g. \cite{mur94}
and \cite{sas96}; because of the complex scattering geometry it
is far less precise than determinations based on Galactic reflection
nebulae, e.g., \cite{wit92}.  Our value depends critically on the 
smoothness of the dust distribution as well as the relative positions
along the line-of-sight of illuminators and scatterers (\cite{wit96},
Gibson \& Nordsieck 1999, in preparation).
Also, as shown in Figure \ref{quad}, 
the true distribution of ultraviolet
luminosity in the LMC contains a non-negligible component due to 
stars not in the Lucke-Hodge associations.  This additional source
of direct, unpolarized, light would dilute the scattered light and
hence reduce $\langle \mathrm P \rangle$ if included in the models.  In such
a model, higher values of $g$ would be required to account for the
observed level of polarization.


\subsection{Individual associations}

Paper I lists nine UV-bright regions which appear
to support scattering halos with the expected centro-symmetric 
pattern; we explore here the degree to which the simple model
is able to reproduce these features of the polarization map.

\subsubsection{NGC 1755, NGC 1711, N 186}

Two open clusters and a modest-sized star-forming region,
these objects show some evidence of scattering halos; however,
they are not included in our models.

\subsubsection{LH 15, N 11, LH 12} 

These bright OB associations support scattering halos.  The
models reproduce them quite strongly.  The northern associations
LH 15 and N11 dominate the local UV radiation field, and are 
well detected in the WISP observations.  LH 12 is far brighter
in the model than is observed, a consequence of the strong
deviation from smoothness of the surrounding dust structures
(see \S6.1).  The observed scattering halo around LH 12 is
quite weak, indicative of the contribution of 
increasing field star density in the southern half of the
WISP field (Paper I).

\subsubsection{LH 4, LH 25, LH 16-17-20}

These regions show very weak evidence of scattering halos in 
the WISP image.  In the models, there is little apparent 
indication of centro-symmetry about these associations.  We 
note that each of these associations lies very close to the
edge of the WISP image.  Moreover, they lie near or within
the scattering halos of brighter associations.  These must
be considered marginal detections.

\subsubsection{B2}

The B2 complex is made up of associations LH 1, 2, 5, and 8;
its large OB star population led us to expect the presence
of a strong centro-symmetric scattering halo.  To the contrary,
the paper I analysis of the WISP observations found
no such halo.  Possible explanations were suggested:
the location of B2 above the plane of the dust layer, location
of B2 within a large H {\small I} hole, or a possible bias 
against detection of scattering halos larger than $\approx$40$\arcmin$.
Our model polarization maps show a centro-symmetric halo around 
B2; however, comparison to the comparable associations N11 and LH 15
showed it to be weaker than the halos of the northern associations.
This is due, in the models, to the extnded 
(non-point-source) size of B2, as well as the contribution of
light from LH 12 and other associations beyond the WISP field of 
view.  In addition, the observations are degraded by the presence 
of high field star density in the southern WISP field, as well
as the existence of a significant hole in the H {\small I}
distribution (S. Kim, private communication).  Part of this
H {\small I} hole corresponds to the H$\alpha$\ supergiant
shell LMC-7 (Meaburn 1980), which coincides with LH 8.
We conclude
that the Paper I non-detection of a scattering
halo around B2 is astrophysical and not due to bias in their
analysis, but the placement of the complex above the plane
of the dust is not required by the observations.

\subsection{Limitations \& Future Work}

This suite of models represents our first attempt to model the
radiation transfer of polarized light through a galaxy from a
large number of discrete sources within a non-uniform dust layer.
These models are able to reproduce the general morphology
of ultraviolet and infrared images of the LMC; taken together
with observed polarization data, they yield astrophysically
interesting constraints on its inclination, and show consistency
with dust properties expected from observations of the Milky Way.

However, the models provide a greatly simplified picture of the
true structure of the LMC.  Future adaptations of the radiation
transfer code will address many of the simplifications; for others,
additional observational material is required in order to refine
our input parameters.  A more advanced version of the Monte Carlo
code (dubbed ``galaxy on a grid''), will allow the specification 
of dust and luminosity density at each individual point in the
model grid; this flexibility will permit the exploration of 
arbitrarily complex geometries in future work.

The most severe drawback of our model is the adoption of a smooth
dust distribution that lacks optically thick clumps.  As noted
by earlier authors (e.g., \cite{wit92a}; \cite{wit96}), the presence
of small-scale, dense knots of absorbing material can greatly alter
the emergent spectral-energy distribution of a galaxy's light.  In
this case, inferred values of $g$ are incorrect, and our ability
to distinguish dust optical properties from the scattering 
geometry is lost (\cite{gib97}, Gibson \& Nordsieck 1999, in preparation).
Since the distribution of small-scale dust
knots in the LMC is unknown, we have little recourse but to
adopt a relatively homogeneous distribution of dust.  The obvious
effect of this approximation is that our models
lack the large polarization fluctuations seen in the observations,
presumably attributable to the presence of the patchy distribution
of optically thick dust clouds.

A related drawback is the unknown line-of-sight distribution
of illuminators relative to the dust.  As shown by \cite{wit92a},
the relative amounts of scattering and absorption depend 
sensitively on where the illuminating sources lie relative to 
the dust.  Along the same lines, we have modelled the supergiant
shells as spheres, whereas evidence suggests they more closely
resemble cylinders.  One obvious effect of this mismatch is 
the presence of a foreground ``haze'' of IR emission above
the shells in our models, which is not present in the IRAS data
(see Figure \ref{quad}).

The dust scale length in our models is highly uncertain; if
we had chosen to adopt the scale length of H {\small I} inferred
from the maps of \cite{kim98}, our derived $r_d$ would have been
some 40\% smaller than the value we used.  This in turn would
have required a higher $\tau_{eq}$ in order to reproduce the 
UV$-$IR color of the diffuse light at the position of the WISP
observations.  Future models, using the galaxy on a grid system,
would be set up to more closely match the complex column-density
variations seen in the \cite{kim98} H {\small I} maps, and therefore
obviate the necessity of parameterization using $r_d$.

Although our scattering medium is unrealistically smooth, our
stellar sources suffer from the opposite problem: we have considered
only emission from large OB associations, ignoring the contributions
of young, massive, open clusters and the field star population.
This yields a model with a very highly clumped luminosity distribution;
because our models ignore the direct stellar contribution to the
diffuse ultraviolet light, they will tend to produce higher levels
of polarization that would otherwise be observed.  For this reason,
our predicted value of $g$ is likely to be skewed towards lower
values than a more complex model would produce.
In future work,
it will be desirable to add a smoothly varying component to the 
starlight, taken for example from the large-scale maps of 
\cite{mau80}.

\subsection{Predictions}

Our models extend over the central $\approx$10$\arcdeg$ $\times$ 10$\arcdeg$
of the Large Magellanic Cloud, roughly 14 times the area covered by the WISP
observations.  This allows us to predict the general pattern of polarization
that might be seen across the face of the LMC at near-ultraviolet wavelengths.
Our models have been tuned to reproduce the optical
depths and polarizations of the WISP field; a test of their
validity and generality would be a comparison of our predictions to future
ultraviolet polarimetric datasets across rest of the Large Cloud.  

\begin{figure*}
\centerline{\hbox{\psfig{figure=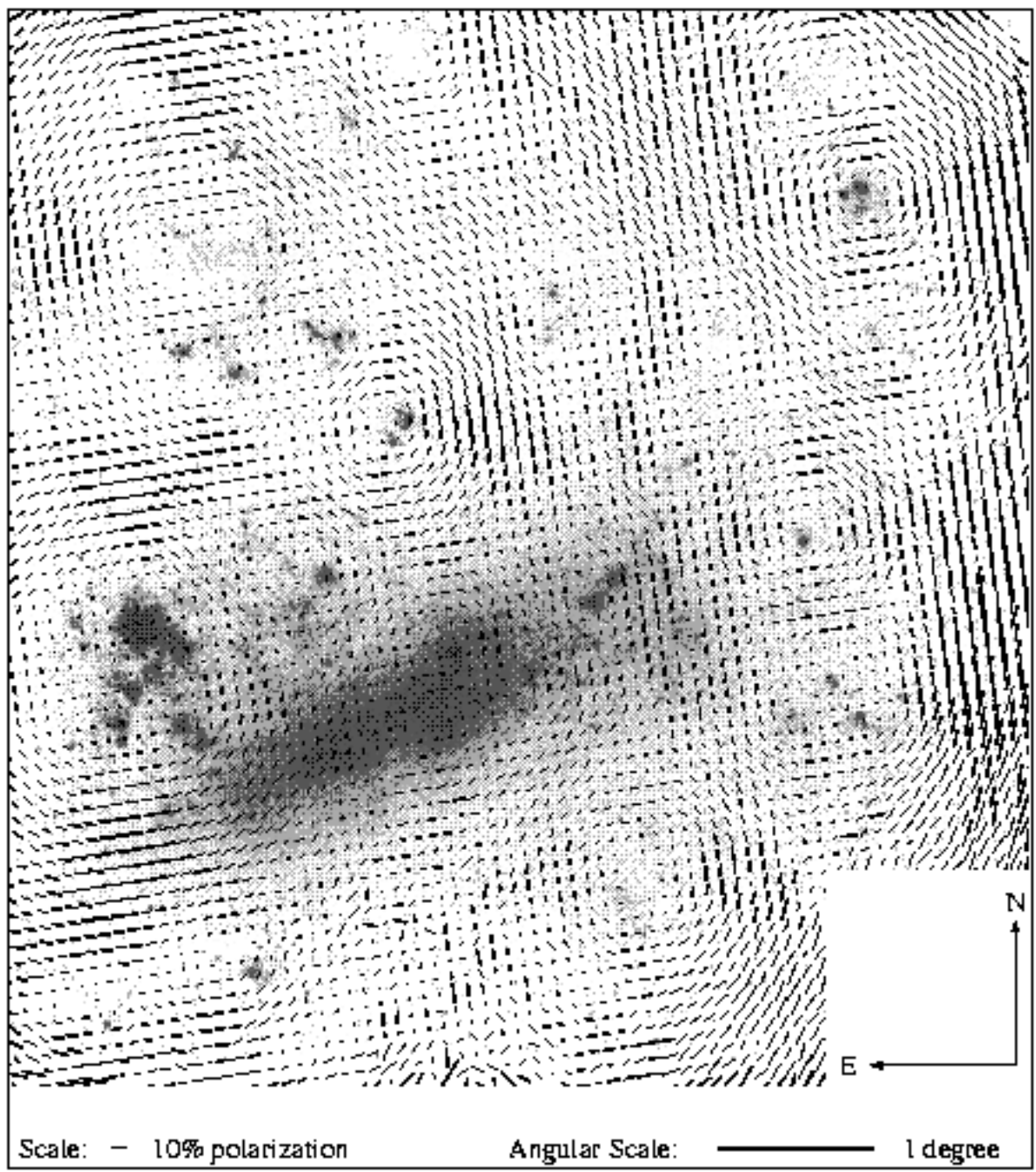}}}
\caption{Polarimetric predictions for the LMC:
our model UV polarization vectors are overlain on an optical image
($\lambda$ $\approx$6600 \AA)
of the galaxy from Sandage (1961).  The WISP field of view is at the
right. North is up, East is to the left.
\label{predict}}
\end{figure*}

In Figure \ref{predict}, we show the 
ultraviolet polarization vectors from our best model
for the entire LMC.  An optical image of the galaxy (Sandage 1961)
is plotted to provide orientation and scale.  If our model is valid,
Figure \ref{predict} should predict the pattern of polarization across the face
of the LMC.  The predictions of the model are most uncertain in the
region of the bar, which contains a high-surface brightness,
intermediate-age population of field stars and numerous young open 
clusters not acounted for in our models (see, e.g., \cite{hod67}).
We comment upon some regions of interest that may be likely future
targets for polarimetric study.

We see that the 30 Dor region
itself shows low levels of polarization, although it is responsible
for much of the scattered light within a kiloparsec or more.  The
bar is expected to show low levels of polarization ($\lesssim$ 5\%),
with most of that due to 30 Dor at the eastern end.  The 
bright H {\small II} region N51 (north of the central bar)
should produce a
centrosymmetric pattern similar to that of N11 (in the WISP field),
although it will be weakened on the eastern and southern sides by
radiation from 30 Dor.  Finally, the southern spiral arm seen
in the H {\small I} maps of \cite{kim98}, beyond the southern limit
of Figure \ref{predict}, is predicted to scatter light from as far north as
30 Doradus, giving rise to faint diffuse light that is polarized
at the 20--30\% level.   
It is expected that these models overpredict
the mean levels of polarization because of the neglected contribution
of direct light from field stars.

\section{Summary}

Using a Monte Carlo radiation transfer code, we have modelled
the observed ultraviolet polarization maps of the LMC obtained
with the WISP instrument.  Our code follows the tracks of 
stellar photons from their origins within OB associations,
through a smoothly distributed exponential dust disk containing
low-density cavities.  By accounting for far-infrared thermal
emission from heated dust, we are able to parameterize the
total amount of dust present by its optical depth.  

Dust-scattered starlight gives rise
to linear polarizations; the magnitude and position angle of
the polarization vectors allow us to derive information regarding
the scattering geometry of stars$+$dust within the LMC and the
optical properties of Magellanic Cloud dust.  We consider three
disk inclinations between 28$\arcdeg$ and 45$\arcdeg$, and
five values for the phase function asymmetry parameter $g$,
between 0.64 and 0.90.  We derive:

$\bullet$ The inclination of the disk of the LMC to the plane
of the sky is 36$\arcdeg^{+2}_{-5}$.  This is in agreement
with other results (\cite{wes97}), but not with a recent 
kinematic determination (\cite{kim98}).  Our determination 
contains a dependence on the dust scale height; we have
assumed $z_d$ = 200 pc, but smaller values would imply larger
inclinations, and vice versa.

$\bullet$ The most likely value for $g$ of 0.70; the uncertainty
in the observations does not permit us to rule out {\it any} 
value for $g$.  Higher signal-to-noise data would sharpen our
estimate, but a more precise method for estimating $g$ from
our models is needed as well.
For $g$ above 0.77, the models produce a higher fraction
of highly polarized ($p$ $\gtrsim$ 20\%) pixels than are observed.
However, the neglect of direct light from field stars probably
leads us to understimate $g$.
Our value $g$ $\approx$ 0.7 is consistent with values derived
from ultraviolet surface photometry of the Galactic reflection
nebula NGC 7023 (\cite{wit92}).

$\bullet$ Our best model predicts that scattered light from 30 Doradus
dominates the eastern side of the LMC's diffuse UV radiation
field; the presence of this ``mini-starburst'' region may be
felt as far south as $-$72$\arcdeg$.  Data for this region would
be of great value in constraining the LMC's structure and geometry.

Our models are the first attempt to replicate the observed
images and polarization maps of the Large Magellanic Cloud.
It is encouraging that our model is able to approximately
match the observations of Paper I for a reasonable set of
input parameters; we do not expect that we have found a 
unique solution for i, $g$, or the geometric parameters
that go into the model.  The modelling procedure should
be applicable to more general problems in the interpretation
of galaxy polarization maps.
However, a large number of simplifying
approximations and assumptions have been made.  The most
serious of these is the smoothness of the adopted model dust
distribution.  Future work will be able to take into account
the complexity of the LMC's H {\small I} distribution, which
is dominated by flocculent spiral arms, supergiant shells,
and small scale filaments (\cite{kim98}).

\acknowledgments

WISP is supported by NASA grant NAG5--647.  K.W. acknowledges
support from NASA's Long Term Space Astrophysics Research 
Program (NAG5--6039).
This modelling project began as a simple exercise in radiation
transfer with Joe Cassinelli in the spring of 1996 and has 
continued to grow from there. A.A.C. would like to thank Joe 
Cassinelli and Jay Gallagher for their patience and encouragement
during the intervening years.   We would like to thank the 
anonymous
referee for suggestions which significantly improved the 
clarity of this paper.

{}

\newpage

\newpage

\appendix

\section{Model Geometry}

Large OB associatons are assumed to 
dominate the diffuse ultraviolet radiation field of the model galaxy.
The source list has been taken from the catalog of Lucke \& Hodge (1970), and
associations are identified by their LH number.  Their radii and fluxes
have been derived for our model purposes in \S2.1.  Our adopted 
source parameters have been listed in Table A1.  Source positions 
are given in terms of the rectilinear model coordinate grid, where
$x$ and $y$ are the east-west and north-south offsets, respectively,
from the optical center of the LMC bar at 
05$\mathrm ^h$ 24$\mathrm ^m$, $-$69$\arcdeg$ 50$\arcmin$ $ $ (B1950.0).
The source radii do not replicate the true physical sizes of the LH
associations, but are the effective radii of circular regions of the
equivalent areas of the associations (see \S2.1).  21 of the associations
lie within the area observed by WISP (Paper I).

The smoothly distributed exponential disk of dust in our models has
been seeded with cavities that approximate the extremely low density
supergiant shells investigated by \cite{mea80}.   The supergiant
shells were identified by H$\alpha$ -emission from their limbs,
and roughly correspond to areas of low H{\small I} column density
(Kim {\it et al.} 1998). While the real supergiant
shells are approximately cylindrical, our code as currently implemented
allows only spherical cavities.  The resulting shells are overlain by 
the high-$z$ tail of the vertical dust distribution and thus fail to precisely
reproduce the infrared morphology of the shells.  The cavity parameters,
with identifications from \cite{mea80}, are given in Table A2.  In 
each case, the optical depth at 2150 \AA $ $ across the diameter of
a cavity has been set to 0.1 (see \S2.2).  Three of the shells are 
contained within the observed WISP field (Paper I).

\begin{deluxetable}{rrrrr|rrrrr|rrrrr}
\tablewidth{0pt}
\tablenum{A1}
\tablecaption{Source parameters for Monte Carlo models.}
\tablehead{
\colhead{LH\tablenotemark{a}} &
\colhead{x\tablenotemark{b}} &
\colhead{y\tablenotemark{b}} &
\colhead{radius\tablenotemark{c}} &
\colhead{flux\tablenotemark{d}} &
\colhead{LH} &
\colhead{x} &
\colhead{y} &
\colhead{rad.} &
\colhead{flux} &
\colhead{LH} &
\colhead{x} &
\colhead{y} &
\colhead{rad.} &
\colhead{flux}}
\startdata
1$^{\star}$ &   2.312 &  -0.250 &     4.05 &   4.91 &
42 &  -0.050 &  -0.300 &     6.00 &  14.45 &
83 &  -1.580 &   3.240 &     3.45 &   5.01 \nl
2$^{\star}$ &   2.250 &  -0.200 &     2.10 &   3.40 &
43 &  -0.100 &   3.750 &     4.35 &   5.75 &
84 &  -1.532 &   2.380 &     4.65 &  26.46 \nl
3$^{\star}$ &   2.467 &   1.732 &     4.35 &   0.87 &
44 &  -0.200 &   0.267 &     6.00 &   0.42 &
85 &  -1.500 &   0.500 &     2.70 &  14.20 \nl
4$^{\star}$ &   2.467 &   2.132 &     4.35 &   6.26 &
45 &  -0.200 &   3.480 &     8.40 &  12.91 &
86 &  -1.500 &   1.820 &     2.55 &   0.79 \nl
5$^{\star}$ &   2.100 &   0.000 &     4.05 &   8.69 &
46 &  -0.250 &  -0.080 &     3.30 &   2.09 &
87 &  -1.450 &  -0.240 &     4.95 &  46.07 \nl
6$^{\star}$ &   2.267 &   1.932 &     4.35 &   3.98 &
47 &  -0.267 &   1.380 &     4.50 &  92.73 &
88 &  -1.532 &   1.732 &     2.10 &   0.46 \nl
7 &   1.700 &  -2.160 &     2.85 &   0.08 &
48 &  -0.280 &   1.425 &     2.85 &   7.37 &
89 &  -1.500 &   0.410 &     5.85 &  93.84 \nl
8$^{\star}$ &   1.920 &  -0.250 &     6.75 &  10.39 &
49 &  -0.350 &   1.250 &     2.85 &   4.87 &
90 &  -1.500 &   0.200 &     2.85 &  19.17 \nl
9$^{\star}$ &   2.240 &   2.600 &     3.00 &  34.88 &
50 &  -0.450 &  -1.932 &     6.75 &   1.10 &
91 &  -1.650 &   2.780 &     2.55 &   0.52 \nl
10$^{\star}$ &   2.230 &   2.700 &     2.85 &   8.71 &
51 &  -0.550 &   1.840 &     2.55 &   2.40 &
92 &  -1.600 &   1.820 &     1.80 &   0.50 \nl
11$^{\star}$ &   1.932 &   0.425 &     3.60 &   0.40 &
52 &  -0.600 &   3.020 &     3.30 &   2.29 &
93 &  -1.580 &   0.020 &     1.80 &   9.77 \nl
12$^{\star}$ &   1.950 &   0.750 &     6.00 &  21.82 &
53 &  -0.650 &   3.150 &     6.30 &   2.75 &
94 &  -1.550 &  -0.080 &     1.80 &  28.18 \nl
13$^{\star}$ &   2.150 &   2.650 &     2.10 &   2.51 &
54 &  -0.637 &   1.820 &     2.55 &  16.60 &
95 &  -1.720 &   2.900 &     3.00 &   2.29 \nl
14$^{\star}$ &   2.100 &   2.732 &     2.25 &   0.12 &
55 &  -0.637 &   1.680 &     4.50 &   1.82 &
96 &  -1.580 &  -0.040 &     7.95 & 151.84 \nl
15$^{\star}$\tablenotemark{\dag} &   2.020 &   3.332 &     5.25 &   3.78 &
56 &  -0.750 &  -2.175 &     4.35 &   0.44 &
97 &  -1.650 &   0.040 &     2.85 &  10.71 \nl
16$^{\star}$ &   1.350 &   0.250 &     2.55 &   1.83 &
57 &  -0.680 &   0.070 &     3.00 &   2.19 &
98 &  -1.620 &  -0.050 &     2.10 &   4.68 \nl
17$^{\star}$ &   1.312 &   0.200 &     2.55 &   0.69 &
58 &  -0.670 &   0.562 &     4.35 &  34.67 &
99 &  -1.620 &   0.220 &     2.85 &   1.91 \nl
18 &   1.250 &  -1.037 &     5.55 &   2.86 &
59 &  -0.760 &  -0.450 &     4.35 &   1.32 &
100 &  -1.720 &   0.280 &     6.00 &  25.12 \nl
19$^{\star}$ &   1.467 &   1.867 &     5.55 &   6.18 &
60 &  -0.780 &   1.852 &     4.35 &  17.72 &
101 &  -1.750 &  -0.100 &     3.60 &  63.73 \nl
20$^{\star}$ &   1.267 &   0.267 &     2.10 &   0.36 &
61 &  -0.780 &   0.400 &     2.55 &  12.59 &
102 &  -1.900 &   1.870 &     3.60 &   1.15 \nl
21$^{\star}$ &   1.380 &   1.460 &     3.60 &   0.79 &
62 &  -0.800 &  -1.160 &     4.80 &   1.00 &
103 &  -1.820 &  -0.250 &     4.20 &  21.06 \nl
22$^{\star}$ &   1.360 &   1.650 &     3.30 &   0.52 &
63 &  -0.800 &   1.900 &     2.70 &  14.07 &
104 &  -1.840 &   0.000 &     3.90 &  23.19 \nl
23 &   1.080 &  -1.560 &     2.85 &   0.25 &
64 &  -0.880 &   0.600 &     6.75 &  15.14 &
105 &  -1.800 &  -0.350 &     2.85 &   0.79 \nl
24 &   1.100 &  -1.300 &     6.90 &  11.71 &
65 &  -1.050 &   2.350 &     2.25 &   1.74 &
106 &  -1.880 &  -0.160 &    14.40 &  52.48 \nl
25$^{\star}$ &   1.250 &   1.120 &     3.30 &   2.00 &
66 &  -1.000 &  -1.620 &     3.30 &   2.05 &
107 &  -1.820 &  -1.800 &     6.90 &   1.20 \nl
26 &   1.050 &  -1.150 &     6.30 &  17.72 &
67 &  -1.050 &   0.125 &     3.30 &   5.87 &
108 &  -1.920 &  -0.250 &     3.00 &   2.51 \nl
27 &   1.000 &   0.160 &     3.30 &   0.24 &
68 &  -1.120 &   0.560 &     0.90 &   0.03 &
109 &  -2.000 &   0.450 &     3.30 &   0.24 \nl
28 &   0.820 &  -1.820 &     5.10 &   1.82 &
69 &  -1.100 &  -1.600 &     4.05 &   7.43 &
110 &  -1.865 &  -1.880 &     3.60 &   0.60 \nl
29 &   0.900 &  -0.667 &     3.60 &   0.36 &
70 &  -1.160 &   1.950 &     2.70 &   2.00 &
111 &  -2.020 &   0.190 &     3.90 &  15.22 \nl
30 &   0.900 &   0.050 &     2.55 &   0.17 &
71 &  -1.160 &   0.750 &     2.55 &   1.51 &
112 &  -2.180 &   1.950 &     2.85 &   0.93 \nl
31 &   0.800 &   0.400 &     4.80 &   9.12 &
72 &  -1.240 &   2.840 &     4.05 &   9.55 &
113 &  -2.050 &   0.320 &     1.80 &   0.26 \nl
32 &   0.900 &   2.080 &     6.30 &   3.31 &
73 &  -1.160 &   0.650 &     2.55 &   0.46 &
114 &  -2.200 &   1.425 &     3.00 &   1.58 \nl
33 &   0.667 &   0.160 &     3.30 &   0.28 &
74 &  -1.200 &   0.150 &     3.90 &   5.35 &
115 &  -2.400 &   2.867 &     6.75 &   3.88 \nl
34 &   0.680 &   1.960 &     4.50 &   2.75 &
75 &  -1.180 &   1.767 &     2.10 &   2.63 &
116 &  -2.380 &   2.020 &     4.80 &   7.24 \nl
35 &   0.440 &   0.000 &     4.95 &  10.96 &
76 &  -1.200 &   1.650 &     5.25 &  39.81 &
117 &  -2.532 &  -0.680 &     3.60 &  15.69 \nl
36 &   0.550 &   1.960 &     2.10 &   1.91 &
77\tablenotemark{\dag} &  -1.420 &   2.350 &     9.60 &  37.84 &
118 &  -2.600 &  -0.732 &     2.85 &   2.49 \nl
37 &   0.562 &   1.900 &     1.80 &   1.51 &
78 &  -1.300 &   1.780 &     2.85 &   3.80 &
119 &  -2.780 &   0.980 &     2.10 &   0.10 \nl
38 &   0.562 &   1.820 &     2.70 &   1.28 &
79 &  -1.300 &   1.850 &     3.90 &   1.91 &
120 &  -2.820 &   1.040 &     4.50 &   0.55 \nl
39 &   0.400 &  -0.160 &     3.90 &   2.51 &
80 &  -1.300 &  -0.532 &     3.60 &   0.38 &
121 &  -3.000 &   1.000 &     7.95 &   1.74 \nl
40 &   0.000 &  -1.820 &     2.10 &   0.13 &
81 &  -1.350 &  -0.300 &     4.95 &  55.39 &
122 &  -3.320 &   0.960 &     4.05 &   0.55 \nl
41 &   0.050 &   0.132 &     6.75 &  50.42 &
82 &  -1.467 &   1.732 &     3.30 &   2.40 &
   &          &          &          &        \nl
\enddata
\tablenotetext{a}{Lucke-Hodge number, see \S2.1.}
\tablenotetext{b}{Offsets in degrees from 5$^{\mathrm{h}}$ 
24$^{\mathrm{m}}$, $-$69$\arcdeg$ 50$\arcmin$\ (B1950.0),
increasing north and west.}
\tablenotetext{c}{Radii in arcmin, see \S2.1.}
\tablenotetext{d}{10$^{-12}$ erg s$^{-1}$ cm$^{-2}$ \AA$^{-1}$}
\tablenotetext{\dag}{Irregular, nonspherical morphology.}
\tablenotetext{\star}{Within observed WISP field.}
\end{deluxetable}

\begin{deluxetable}{crrr}
\tablewidth{0pt}
\tablenum{A2}
\tablecaption{Cavity parameters for Monte Carlo models.} 
\tablehead{
\colhead{Shell\tablenotemark{a}} &
\colhead{x\tablenotemark{b}} &
\colhead{y\tablenotemark{b}} &
\colhead{radius\tablenotemark{c}} }
\startdata
LMC-1$^{\star}$ & 2.025 & 3.325 & 24.0 \nl
LMC-2 & -2.225 & -0.475 & 30.0 \nl
LMC-3 & -1.075 & 0.350 & 34.5 \nl
LMC-4 & -1.100 & 2.375 & 40.5 \nl
LMC-5 & -0.600 & 2.900 & 27.0 \nl
LMC-6$^{\star}$ & 1.950 & 0.575 & 21.0 \nl
LMC-7$^{\star}$ & 2.125 & -0.250 & 27.0 \nl
LMC-8 & 1.075 & -1.225 & 30.0 \nl
LMC-9 & -0.725 & -1.775 & 30.0 \nl
\enddata
\tablenotetext{a}{From Meaburn (1980), see \S2.2.}
\tablenotetext{b}{Offsets in degrees from 5$^{\mathrm{h}}$
24$^{\mathrm{m}}$, $-$69$\arcdeg$ 50$\arcmin$\ (B1950.0),
increasing north and west.}
\tablenotetext{c}{Radii in arcminutes, see \S2.2.}
\tablenotetext{\star}{Within observed WISP field.}
\end{deluxetable}

\end{document}